\documentclass[showpacs,preprintnumbers,amsmath,amssymb]{revtex4}

\usepackage{graphicx}
\usepackage{dcolumn}
\usepackage{bm}

\begin{document}

\title{Theoretical and experimental $\alpha$ decay half-lives of the heaviest odd-Z elements
and general predictions}

\author{H.F. Zhang$^{1}$}

\author{G. Royer$^{2}$}

\affiliation{\footnotesize$^1$ School of Nuclear Science and
Technology, Lanzhou University, Lanzhou 730000, P. R. China}

\affiliation{\footnotesize $^2$Laboratoire Subatech, UMR:
IN2P3/CNRS-Universit\'e-Ecole des Mines, 4 rue A. Kastler,
 44307 Nantes Cedex 03, France}

\date{\today}

\begin{abstract}
Theoretical $\alpha$-decay half-lives of the heaviest odd-Z nuclei 
 are calculated using the experimental $Q_{\alpha}$ value. The barriers in the
quasi-molecular shape path is determined within a Generalized Liquid Drop Model (GLDM) and the 
WKB approximation is used.
 The results are compared with calculations
using the Density-Dependent M3Y (DDM3Y) effective interaction and
the Viola-Seaborg-Sobiczewski (VSS) formulae. The calculations
provide consistent estimates for the half-lives of the 
$\alpha$ decay chains of these superheavy elements. The
experimental data stand between the GLDM calculations and VSS
ones in the most time. Predictions are provided for the
$\alpha$ decay half-lives of other superheavy nuclei within the GLDM and VSS
approaches using the extrapolated Audi's recent $Q_{\alpha}$, which may be
used for future experimental assignment and identification.

\end{abstract}
\pacs{27.90.+b, 23.60.+e, 21.10.Tg}

\maketitle

The possibility to synthesize superheavy elements by cold or warm
fusion reactions \cite{Ho00,Ar00,Og99} or using radioactive ion
beams has renewed interest in investigating the fusion barriers.
The only observed decay mode of these heaviest systems is the
$\alpha$ emission, and an accurate description of the $\alpha$
decay is required. The pure Coulomb barrier sharply peaked at the
touching point alone does not allow to determine correctly the
fusion cross sections and the partial $\alpha$ decay half-lives.
In the fusion path, the nucleon-nucleon forces act before the
formation of a neck between the two quasispherical colliding ions
and a proximity energy term must be added in the usual development
of the liquid-drop model \cite{Bl77}. It is highly probable that the $\alpha$ decay
takes place also in this fusion-like deformation valley where the
one-body shape keeps quasi-spherical ends while the transition
between one and two-body configurations corresponds to two
spherical nuclei in contact. Consequently, the proximity energy
term plays also a main role to correctly describe the $\alpha$
decay barrier. The generalized liquid drop model (GLDM) which
includes such a proximity energy term has allowed to describe the
fusion \cite{Roy85} , fission \cite{royzb02}, light nucleus
\cite{rm01} and $\alpha$ emission \cite{Roy00} processes. The formation and alpha decay of superheavy elements 
have been investigated \cite{rogh02} in taking into account the experimental $Q_{\alpha}$ value
or the value provided by the Thomas-Fermi model \cite{ms96}. The heaviest even-Z  
nuclei have been studied \cite{Zh06} using the $Q_{\alpha}$ value obtained experimentally or given by the FRDM \cite{mn97}.

Recently, isotopes of the element 115 have been synthesized
\cite{Og04} and the observed decays reveal that the dominant decay
mode is the $\alpha$ emission. These new experimental observations
of Z=115 have already attracted a lot of theoretical studies 
\cite{Re03,Ge03,Da04,Sh05,Ga05,Zh05,Ba05,Ba04}. Most of the
earlier investigations have been devoted to the description of the
ground-state properties of superheavy nuclei, we focus on
calculating their half-lives following the first work for even-Z nuclei \cite{Zh06}. 
In Ref. \cite{Ba04}, the $\alpha$ decay half-lives of Z =
115 isotopes are calculated with the microscopic density-dependent
M3Y (DDM3Y) interaction, and the results are well consistent with the
experimental data. The purpose of this work is to determine
the partial $\alpha$ decay half-lives of these superheavy elements
within the macroscopic GLDM from the experimental $Q_{\alpha}$
values using the WKB approximation and to compare with the
experimental data and the calculations of DDM3Y effective
interaction \cite{Ba04} and the Viola-Seaborg formulae with
Sobiczewski constants (VSS) \cite{So89}. Finally predictions
within the GLDM and VSS formulae are given for the partial
$\alpha$ decay half-lives of the superheavy nuclei using the
recent $Q_{\alpha}$ decay energies of Audi et al. \cite{Au03}.

The GLDM energy  
 is widely explained in \cite{Zh06} and not recalled here.
The half-life of the parent nucleus decaying via $\alpha$ emission
is calculated using the WKB barrier penetration probability. In such a
unified fission model, the decay constant of the $\alpha$ emitter
is simply defined as
    $\lambda=\nu_{0}P$
where the assault frequency $\nu_{0}$ has been taken as
    $\nu_{0}=10^{19}s^{-1}$, P being the barrier penetrability.

The $\alpha$ decay half-lives  of the recently produced odd-Z
superheavy nuclei calculated with the three approaches and using the experimental
$Q_{\alpha}$ values and without considering the rotational
contribution are presented in Table 1. The $Q_{\alpha}$ values given in \cite{Au03}
 are obtained by extrapolation. Within the GLDM the quantitative agreement
with experimental data is visible. The experimental half-lives are
reproduced well in six cases ( $^{288}$115, $^{284}$113,
$^{272}$107, $^{287}$115, $^{283}$113, $^{275}$109) out of nine
nuclei along the decay chains of $^{288}$115 and $^{287}$115. Two
results ( $^{280}$111, $^{276}$109) are underestimated about four
to five times possibly because the centrifugal barrier required
for the spin-parity conservation could not be taken into account
due to non availability of the spin-parities of the decay chain
nuclei. On the whole, the results agree well with the
experimental data indicating that a GLDM taking account the
proximity effects, the mass asymmetry, and an accurate nuclear
radius is sufficient to reproduce the $\alpha$ decay potential
barriers when the experimental $Q_{\alpha}$ value is known. The
results obtained with the DDM3Y interaction agree with the
experimental data as the GLDM predictions and largely better than
the VSS calculations. This shows that a double folding potential
obtained using M3Y \cite{Be77} effective interaction supplemented
by a zero-range potential for the single-nucleon exchange is very
appropriate because its microscopic nature includes many nuclear
features, in particular a potential energy surface is inherently
embedded in this description. This double agreement shows that the
experimental data themselves seem to be consistent. For most
nuclei the predictions of the VSS model largely overestimate the
half lives. The blocking effect is probably treated too roughly.

\begin{table*}[h]
\label{table1} \caption{Comparison between experimental $\alpha$
decay half-lives \cite{Og04} and results obtained
 with the GLDM, the DDM3Y effective
interaction \cite{Ba04} and the VSS formulae. }
\begin{ruledtabular}
\begin{tabular}{lllllllllll}
Parent& Expt.& $[23]$& Expt.& DDM3Y &GLDM & GLDM & VSS & VSS  \\
Nuclei&$Q$[MeV]&$Q$[MeV]&$T_{1/2}$&$T_{1/2}(Q_{ex})$&$T_{1/2}(Q_{ex})$&$T_{1/2}(Q_{Audi})$&$T_{1/2}(Q_{ex})$&$T_{1/2}(Q_{Audi})$\\
\hline
&&&&&&& \\
  $^{288}$115  &10.61 (6)&      & 87 $^{+105}_{-30}$ ms    &    409            ms  &94.7$^{+41.9}_{-28.9}$ ms  &        &997$^{+442}_{-303}$  ms    \\
&&&&&&& \\
  $^{284}$113  &10.15 (6)&10.25 & 0.48$^{+0.58}_{-0.17}$ s &1.55$^{+0.72}_{-0.48}$s&0.43$^{+0.21}_{-0.13}$ s   & 0.23 s & 4.13$^{+1.94}_{-1.31}$ s & 2.19  s\\
&&&&&&& \\
  $^{280}$111  & 9.87 (6)& 9.98 &  3.6 $^{+4.3}_{-1.3}$  s &1.9$^{+0.9}_{-0.6}$ s  &0.69$^{+0.33}_{-0.23}$ s   & 0.34  s& 5.70$^{+2.74}_{-1.84}$ s & 2.79  s\\
&&&&&&& \\
  $^{276}$109  & 9.85 (6)& 9.80 & 0.72$^{+0.87}_{-0.25}$ s &0.45$^{+0.23}_{-0.14}$s&0.19$^{+0.08}_{-0.06}$ s   & 0.26  s& 1.44$^{+0.68}_{-0.46}$ s & 1.99  s\\
&&&&&&& \\
  $^{272}$107  & 9.15 (6)& 9.30 &  9.8$^{+11.7}_{-3.5}$  s &10.1$^{+5.4}_{-3.4}$ s&5.12$^{+3.19}_{-1.58}$  s   & 1.89  s& 33.8$^{+17.9}_{-11.6}$ s & 11.91  s \\
&&&&&&& \\
  $^{287}$115  & 10.74 (9)&     &   32$^{+155}_{-14}$ ms   &   49  ms              &46.0$^{+33.1}_{-19.1}$ ms  &        & 207$^{+149}_{-85}$  ms   & \\
&&&&&&& \\
  $^{283}$113  & 10.26 (9)&10.60 &100$^{+490}_{-45}$ ms    &201.6$^{+164.9}_{-84.7}$ms&222$^{+172}_{-96}$ ms   & 27.1 ms& 937$^{+719}_{-402}$ s    & 116.7  ms\\
&&&&&&& \\
  $^{279}$111  & 10.52(16)&10.45 &170$^{+810}_{-80}$ ms    &9.6$^{+14.8}_{-5.7}$ ms&12.4$^{+19.9}_{-7.6}$ ms   & 18.8 ms& 45.3$^{+73.1}_{-27.6}$ms & 68.8  ms\\
&&&&&&& \\
  $^{275}$109  & 10.48 (9)&10.12 & 9.7$^{+46}_{-4.4}$  ms  &2.75$^{+1.85}_{-1.09}$ms&4.0$^{+2.8}_{-1.6}$ ms    & 35.2 ms& 13.7$^{+9.6}_{-5.6}$ ms  & 119.5  ms\\
\end{tabular}
\end{ruledtabular}
\end{table*}

One can also find that all calculated half-lives of the 
$^{279}$111 nucleus are smaller than the experimental ones in table 1. If
the contribution of centrifugal barrier is included, the
theoretical results will close the experimental data. On the other
hand, it is expected that great deviations of a few superheavy
nuclei between the data and model may be eliminated by further
improvements on the precision of measurements.

Another noticeable point is that the experimental $\alpha$ decay
half-lives are between the close theoretical values given by the
GLDM and the ones derived from the VSS formulae. Thus predictions
of the $\alpha$ decay half lives with the GLDM and VSS formulae
are possible as long as we know the right $\alpha$ decay
energies. The ones derived from Audi's recent
publication \cite{Au03} are very close to the experimental data.
The most deviation is not more than 0.5 MeV, which is a valuable
result for studying correctly the half-lives. The
calculations using the $\alpha$ decay energies of Ref.\cite{Au03}
for the nuclei of the $^{288}$115 and $^{287}$115 decay chains
by the GLDM and VSS formulae are reasonably consistent with the experimental data. 
The experimental data stand between the
calculations of the GLDM and the results of VSS in six cases for the
seven nuclei when experimental uncertainty in the Q value is
considered. Thus, predictions of the half-lives of superheavy
nuclei with the GLDM and VSS formulae are provided for a
large number of superheavy elements in Table 2 using the extrapolated Q$_{\alpha}$ values given by \cite{Au03} or the experimental data indicated by an asterisk. They are an improvement relatively to the values previously given 
in \cite{rogh02} since these extrapolated Q$_{\alpha}$ values are in better agreement with the experimental data 
than the ones proposed in \cite{ms96}. It may be useful for the future
experimental assignment and identification.

\begin{table*}[h]
\caption {\label{table2}Predicted $\alpha$-decay half-lives using
the GLDM and the VSS formulae, the $\alpha$ decay energies are
taken from the extrapolated data of Audi et al. \cite{Au03} or the experimental data indicated by an asterisk.}
\begin{ruledtabular}
\begin{tabular}{lllllllllllll}
Nuclei &Q [ MeV ]&$T^{GLDM}_{1/2} $&$T^{VSS}_{1/2} $&Nuclei &Q [ MeV ]&$T^{GLDM}_{1/2} $&$T^{VSS}_{1/2} $&Nuclei &Q [ MeV ]&$T^{GLDM}_{1/2} $&$T^{VSS}_{1/2} $\\
\hline
&&&\\
$^{293}$118&12.30& 77$\mu$s &592$\mu$s &$^{292}$117&11.60& 1.30 ms & 13.33 ms &$^{292}$116&10.71& 94.6 ms & 84.7 ms\\
$^{291}$117&11.90& 0.29  ms &  1.23 ms &$^{291}$116&11.00& 17.7 ms & 176 ms   &$^{291}$115&10.00&  4.33 s & 21.9 s \\
$^{290}$116&11.30& 3.36  ms &  2.75 ms &$^{290}$115&10.30& 0.62 s  & 6.86 s   &$^{289}$116&11.70& 0.43 ms & 3.63 ms\\
$^{289}$115&10.60& 97.4 ms  &  482 ms  &$^{289}$114& 9.85& 5.81 s  & 55.72 s  &$^{288}$115&11.00& 9.41 ms & 99.1 ms\\
$^{288}$114& 9.97&  2.67 s  &  2.15 s  &$^{287}$115&11.30& 1.92 ms &  8.29 ms &$^{287}$114&10.44& 0.136 s & 1.24 s\\
$^{287}$113& 9.34& 102 s    &  461 s   &$^{286}$114&10.70& 30 ms   &  22 ms   &$^{286}$113& 9.68& 9.44 s  & 92.5 s\\
$^{285}$114&11.00& 5.1 ms   &  44.6 ms &$^{285}$113&10.02& 0.99 s  &  4.35 s  &$^{285}$112& 8.79& 49.97 m & 425 m \\
$^{284}$113&10.25& 0.23 s   &  2.19 s  &$^{284}$112& 9.30& 64.7 s  &  47.3 s  &$^{283}$113&10.60& 27.1 ms & 116.7ms\\
$^{283}$112& 9.62& 6.93 s   &  58.09 s &$^{283}$111& 8.96& 6.01 m  & 25.73 m  &$^{282}$112& 9.96& 0.772 s & 0.516 s\\
$^{282}$111& 9.38& 18.6 s   & 158.4 s  &$^{281}$112&10.28& 0.102 s & 0.786 s  &$^{281}$111& 9.64& 3.12  s & 11.96 s\\
$^{281}$110& 8.96& 3.05 m   & 22.47 m  &$^{280}$112&10.62& 13.3 ms & 8.62 ms  &$^{280}$111& 9.98& 0.335 s & 2.79 s \\
$^{280}$110& 9.30& 15.5 s   & 9.76 s   &$^{279}$112&10.96& 2.06 ms & 14.1 ms  &$^{279}$111&10.45& 18.8 ms & 68.8 ms\\
$^{279}$110& 9.60& 2.02 s   & 14.3  s  &$^{279}$109& 8.70& 10.35 m & 36.32 m  &$^{278}$112&11.38& 0.223ms & 0.121ms\\
$^{278}$111&10.72& 3.89 ms  & 30.9 ms  &$^{278}$110&10.00& 148.5ms & 89.8 ms  &$^{278}$109& 9.10&   31 s  & 240  s \\
$^{277}$112&11.62& 0.069 ms & 0.402ms  &$^{277}$111&11.18& 0.323 ms& 1.073ms  &$^{277}$110&10.30& 23.1 ms & 162 ms \\
$^{277}$109& 9.50& 1.89  s  & 6.61 s   &$^{277}$108& 8.40& 49.7 m  & 330.3 m  &$^{276}$111&11.32& 0.157 ms& 1.11ms \\
$^{276}$110&10.60& 4.03 ms  & 2.35 ms  &$^{276}$109& 9.80& 0.26  s & 1.99 s   &$^{276}$108& 8.80& 131 s   & 75 s  \\
$^{275}$111&11.55&51.5$\mu$s&152$\mu$s &$^{275}$110&11.10& 0.26 ms & 1.65 ms  &$^{275}$109&10.12& 35.2 ms & 119.5 ms\\
$^{275}$108& 9.20& 7.13 s   & 47.2 s  &$^{274}$111&11.60&41.4$\mu$s&258$\mu$s &$^{274}$110&11.40&55.5$\mu$s&28.7$\mu$s\\
$^{274}$109&10.50& 3.67 ms  & 26.8ms  &$^{274}$108& 9.50& 0.92 s   & 0.51 s   &$^{274}$107& 8.50& 9.94 m  & 70.98 m\\
$^{273}$111&11.20& 0.33 ms  & 0.96 ms &$^{273}$110&11.37&0.067 ms  & 0.39 ms  &$^{273}$109&10.82& 0.61 ms & 1.96 ms\\
$^{273}$108& 9.90& 69.4 ms  & 441.6 ms&$^{273}$107& 8.90& 28.8 s   & 92.8 s   &$^{272}$111&11.44& 0.11 ms & 0.59 ms\\
$^{272}$110&10.76& 1.97 ms  & 0.94 ms &$^{272}$109&10.60& 2.34 ms  & 15.02 ms &$^{272}$108&10.10& 21.7 ms & 10.9 ms\\
$^{272}$107& 9.30& 1.89 s   & 11.91 s &$^{272}$106& 8.30& 24.9 m   & 11.4 m   &$^{271}$110&10.87& 1.12 ms & 5.86 ms\\
$^{271}$109&10.14& 37.5 ms  & 105.6 ms&$^{271}$108& 9.90& 79.2 ms  & 441.7 ms &$^{271}$107& 9.50& 0.499 s & 1.40 s \\
$^{271}$106& 9.20& 1.74 s   & 16.78 s &$^{270}$110&11.20& 0.199 ms & 0.083 ms &$^{270}$109&10.35& 10.7 ms & 65 ms  \\
$^{270}$108& 9.30& 4.48 s   & 2.02 s  &$^{270}$107& 9.30& 2.0 s    & 11.9 s   &$^{270}$106& 9.10& 3.59 s  & 1.66 s\\
$^{270}$105& 8.20& 24.38 m  & 140.53 m&$^{269}$110&11.58& 30$\mu$s & 132$\mu$s&$^{269}$109&10.53& 3.75 ms & 10.25 ms\\
$^{269}$108& 9.63& 0.48 s   & 2.52 s  &$^{269}$107& 8.84& 55.9 s   & 144.5 s  &$^{269}$106& 8.80& 32.5 s  & 167.9 s\\
$^{269}$105& 8.40& 4.96 m   & 12.93 m &$^{268}$110&11.92& 6.3$\mu$s& 2.1$\mu$s&$^{268}$109&10.73& 1.28 ms & 7.15 ms \\
$^{268}$108& 9.90& 85.7 ms  & 37.7 ms &$^{268}$107& 9.08& 9.86 s   & 55.5 s   &$^{268}$106& 8.40& 12.1 m  & 5.1 m \\
$^{268}$105& 8.20& 25.4 m   & 140.5 m &$^{268}$104& 8.10& 23.8 m   & 10.2 m   &$^{267}$110&12.28&1.3$\mu$s& 4.4$\mu$s\\
$^{267}$109&10.87& 0.61 ms  & 1.49 ms &$^{267}$108&10.12& 22.1 ms  & 112.5 ms &$^{267}$107& 9.37& 1.33 s  & 3.36 s \\
$^{267}$106& 8.64& 1.9 m    & 9.3 m   &$^{267}$105& 7.90& 330 m    & 787 m    &$^{267}$104& 7.80& 315  m  & 1494 m \\
$^{266}$109&10.996& 0.32 ms & 1.63 ms &$^{266}$108&10.336& 6.26 ms & 2.64 ms  &$^{266}$107& 9.55& 0.41 s  & 2.21 s\\
$^{266}$106& 8.88& 19.3 s   & 8.02 s  &$^{266}$105& 8.19& 29.0 m   & 152.5 m  &$^{266}$104& 7.50& 81.47 h & 31.30 h\\
$^{265}$109&11.07& 0.223 ms & 0.498 ms&$^{265}$108&10.59& 1.47 ms  & 7.00 ms  &$^{265}$107& 9.77&  99.7ms & 241 ms\\
$^{265}$106& 9.08& 4.7 s    & 22.2 s  &$^{265}$105& 8.49& 2.70 m   & 6.43 m   &$^{265}$104& 7.78& 6.58 h  & 29.65 h\\
$^{264}$108&10.59& 1.58 ms  & 0.60 ms &$^{264}$107& 9.97& 29.9 ms  & 151 ms   &$^{264}$106& 9.21& 1.99 s  & 0.77 s \\
$^{264}$105& 8.66& 46.1 s   & 232 s   &$^{264}$104& 8.14& 19.2 m   & 7.36 m   &$^{263}$108&10.67& 1.03 ms & 4.45 ms \\
$^{263}$107&10.08& 15.5 ms  & 34.9 ms &$^{263}$106& 9.39& 0.60 s   & 2.64 s   &$^{263}$105& 9.01& 3.65 s  & 8.27 s \\
$^{263}$104& 8.49& 72.7 s   & 324.7 s &$^{262}$107&10.30& 4.42 ms  & 20.5 ms  &$^{262}$106& 9.60&160.4 ms & 56.7 ms\\
$^{262}$105& 9.01& 4.06 s   & 18.2 s  &$^{262}$104& 8.49& 82.6 s   & 27.9 s   &$^{261}$107&10.56& 1.04 ms & 2.07 ms \\
$^{261}$106& 9.80& 44.8 ms  & 183.9 ms&$^{261}$105& 9.22& 0.96 s   & 1.92 s   &$^{261}$104& 8.65& 25.0 s  & 97.2 s\\
$^{260}$107&10.47& 1.77 ms  & 7.62 ms &$^{260}$106& 9.92& 21.9 ms  & 7.48 ms  &$^{260}$105& 9.38& 0.33 s  & 1.44 s \\
$^{260}$104& 8.90& 4.09 s   & 1.35 s  &$^{259}$106& 9.83& 39.4 ms  & 152.3 ms &$^{259}$105& 9.62& 69.0 ms & 136.7 ms\\
$^{259}$104& 9.12& 0.89 s   & 3.38 s  &$^{258}$106& 9.67& 114 ms   & 36 ms    &$^{258}$105& 9.48& 0.18 s  & 0.74 ms \\
$^{258}$104& 9.25& 380 ms   & 120 ms  &$^{257}$105& 9.23& 1.0 s    & 1.8  s   &$^{257}$104& 9.04& 1.66 s  & 5.88 s \\
$^{256}$105& 9.46& 230 ms   & 848 ms  &$^{256}$104& 8.93& 3.78 s   & 1.09 s   &$^{255}$105& 9.72& 42.9 ms & 72.4 ms \\
$^{255}$104&9.058& 1.57 s   & 5.19 s  &$^{254}$104& 9.38& 181 ms   & 50.5 ms  &$^{253}$104& 9.55& 63.1 ms & 195.0 ms \\
\end{tabular}
\end{ruledtabular}
\end{table*}

In conclusion, the half-lives for $\alpha$-radioactivity have been
analyzed in the quasimolecular shape path within a Generalized
Liquid Drop Model including the proximity effects between nucleons
and the mass and charge asymmetry. The results are in agreement with
the experimental data for the alpha decay half-lives along the decay chains of the 
Z=115 isotopes and close to the ones derived from the DDM3Y
effective interaction. The experimental $\alpha$ decay half lives
stand between the GLDM calculations and VSS formulae results and
the $\alpha$ decay half-lives of some superheavy nuclei have been
presented within the GLDM and VSS approaches and Q$_{\alpha}$
adopted from the Audi's recent extrapolated data.

H.F. Zhang is thankful to Prof. U. Lombardo, Junqing Li, Wei Zuo,
Baoqiu Chen and Zhong-yu Ma for valuable discussions. H.F. Zhang
also would like to express his gratitude for the kind hospitality
and excellent working conditions at INFN-LNS. Partial support from
the Natural Science Foundation of China (NSFC) under Grant No.
10505016, 10235020, 10235030, 10575119, and the European
Commission under the Grant CN/ASIA-LINK/008(94791).



\begin{thebibliography}{99}

\bibitem{Ho00} S. Hofmann and G. M\"unzenberg, Rev. Mod. Phys. {\bf 72 } 733 (2000).

\bibitem{Ar00} P. Armbruster, Eur. Phys. J. A {\bf 7 } 23 (2000).

\bibitem{Og99} Yu. Ts. Oganessian {\it et al.},  Phys. Rev. Lett. {\bf 83} 3154 (1999) ;
Eur. Phys. J. A {\bf 5}, 63 (1999) 63 ; Nature {\bf 400} 242
(1999).

\bibitem{Bl77} J. Blocki, J. Randrup, W.J. Swiatecki, and C.F. Tsang, Ann. Phys. {\bf 105} (1977) 427.

\bibitem{Roy85} G. Royer and B. Remaud, Nucl. Phys. A {\bf 444}, 477 (1985).

\bibitem{royzb02} G. Royer and K. Zbiri, Nucl. Phys. A {\bf 697}, 479 (2002).

\bibitem{rm01} G. Royer and R. Moustabchir, Nucl. Phys. A {\bf 683}, 182 (2001).

\bibitem{Roy00} G. Royer, J. Phys. G {\bf 26}, 1149 (2000).


\bibitem{rogh02} G. Royer and R. A. Gherghescu, Nucl. Phys. A {\bf 699}, 479 (2002).

\bibitem{ms96} W. D. Myers and W. J. Swiatecki, Nucl. Phys. A {\bf 601}, 141 (1996).

\bibitem{Zh06} H.F. Zhang, W. Zuo, J.Q. Li, and G. Royer, Phys. Rev. C {\bf 74}, 017304 (2006).

\bibitem{mn97} P. M\"oller, J. R. Nix, and K.-L. Kratz, At. Data Nucl. Data Tables {\bf 66}, 131 (1997).

\bibitem{Og04} Yu. Ts. Oganessian {\it et al.}, Phys. Rev. C {\bf 69}, 021601(R) (2004); Phys. Rev. C {\bf 69}, 029902(E) (2004); Phys. Rev. C {\bf 72}, 034611 (2005).

\bibitem{Re03} Zhongzhou Ren, Ding-Han Chen, Fei Tai, H. Y. Zhang, and W. Q.
               Shen, Phys. Rev. C {\bf 67}, 064312 (2003).

\bibitem{Ge03} L. S. Geng, H. Toki, and J. Meng, Phys. Rev. C {\bf 68}, 061303(R) (2003).

\bibitem{Da04} Sankha Das and G. Gangopadhyay, J. Phys. G, {\bf 30}, 957 (2004).

\bibitem{Sh05} M. M. Sharma, A. R. Farhan, and G. M\"unzenberg, Phys. Rev. C {\bf 71}, 054310 (2005).


\bibitem{Ga05} Y.K. Gambhir, A. Bhagwat, and M. Gupta, Annals of Physics, {\bf 320}, 429 (2005).

\bibitem{Zh05} H.F. Zhang, J.Q. Li, W. Zuo, Z.Y. Ma, B.Q. Chen, and Im Soojae, Phys. Rev. C {\bf 71 }, 054312 (2005).

\bibitem{Ba05} E Baldini-Neto, B V Carlson, and D Hirata, J. Phys. G, {\bf 32}, 655 (2006).

\bibitem{Ba04} D. N. Basu, J. Phys. G, {\bf 30}, B35 (2004);
               P. Roy Chowdhury, D. N. Basu, and C. Samanta, Phys. Rec. C {\bf 75 } 047306 (2007).

\bibitem{So89} V. E. Viola and G. T. Seaborg, J. Inorg. Nucl. Chem. 28, 741 (1966); A. Sobiczewski, Z. Patyk, and S. Cwiok, Phys. Lett. B {\bf 224}, 1 (1989).

\bibitem{Au03} G. Audi, A.H. Wapstra, and C. Thibault, Nucl. Phys. {\bf A 729}, 337 (2003).

\bibitem{Be77} G. Bertsch, J. Borysowicz, H. McManus, and W.G. Love, Nucl. Phys. A {\bf 284}, 399 (1977).






\end{thebibliography}
\end{document}